\documentclass{PoS}

\title{Polarised correlators at finite temperature}

\ShortTitle{Polarised correlators at finite temperature}

\author{G.~Aarts, \speaker{C.R.~Allton}, S.J.~Hands \\
Department of Physics, Swansea University, Swansea, U.K.}
\author{J.~Foley \\
Carnegie Mellon University, Pittsburgh, USA}
\author{S.~Kim \\
Sejong University, Seoul, Korea}

\abstract{ QCD undergoes a deconfining transition at high temperature
  to a ``quark-gluon plasma'' phase where hadrons may become unbound. In
  this work, meson correlation functions at non-zero momentum are
  studied both in the confined and plasma phases using the Maximum
  Entropy Method. In particular, both the longitudinal and transverse
  modes of the vector correlation functions are considered. 
  Only in the case of light quarks in the plasma phase, we find 
  that both longitudinal and transverse spectral functions 
  have a non-zero intercept at zero energy.  }

\FullConference{XXVIIIth International Symposium on Lattice Field Theory\\
                June 14-19, 2010\\
                Villasimius, Sardegna, Italia}

\begin{document}



\section{Introduction}

There has been considerable interest in the quark-gluon plasma phase
of QCD over recent years. On the experimental side, experiments at
RHIC have led to an increased understanding of this phase of matter.
This has been matched by a similarly intense theoretical understanding
using e.g.\ hydrodynamic models and lattice simulations.  Despite this
considerable work, there remains a great deal of uncertainty over many
properties of the quark-gluon plasma. In particular, the Particle
Data Book does not contain a single entry for this phase
\cite{Amsler:2008zzb}.

This work aims to extend our knowledge of the physics of QCD in the plasma 
phase by studying the properties of mesonic spectral functions, 
$\rho(\omega)$, both below and above the deconfining temperature, $T_c$. 
The zero energy limit, $\omega \rightarrow 0$, of spectral functions gives 
information on hydrodynamic structure and transport coefficients; we will 
be particularly interested in the conductivity, $\sigma$, and the related 
diffusivity, $D=\sigma/\Xi$, where $\Xi$ is the charge susceptibility. 
These can be obtained from vector spectral functions. In this paper, we 
outline an extension of our earlier work \cite{conductivity} to the case 
of non-zero momentum \cite{Aarts:2006wt}.  This allows us to study the 
longitudinal and transverse components in the vector channel which can, in 
principle, unlock interesting information about transport and 
hydrodynamics \cite{Hong:2010at}.
 For more details on transport coefficients and lattice QCD, we refer 
to the reviews \cite{Aarts:2007va,Meyer:2008sn}.



\section{Lattice and Fitting Details}

We begin by defining the usual spectral function in terms of the
Euclidean two-point function, $G(t,\vec p)$,
\[
G(t,\vec{p}) = \int \frac{d\omega}{2\pi}\,\rho(\omega,\vec{p}) 
K(t,\omega),
\]
where $\vec{p}$ denotes the momentum. 
The kernel is given by
\[
K(t,\omega) = \frac{\cosh[\omega(t-1/(2T))]}{\sinh[\omega/(2T)]},
\]
where $T=1/(aN_t)$.
We note that the extraction of a spectral density from a lattice
correlator is an ill-posed problem, since the correlator, $G(t)$, is
known at only ${\cal O}(10)$ time points, whereas the spectral
function, $\rho(\omega)$, is, in principle, a continuous function.
The usual solution of this problem implements the Bayesian analysis
technique of the Maximum Entropy Method (MEM) \cite{hatsuda}.

At finite temperature the kernel is singular and independent of 
euclidean time in the limit that $\omega\to 0$ \cite{Aarts:2002cc}:
\[
\lim_{\omega\rightarrow 0} K(\omega,t) = \frac{2T}{\omega} +
{\cal O}(\omega).
\]
In Ref.\ \cite{conductivity} we uncovered that this singularity affects 
the reliability of the MEM procedure near $\omega = 0$. Fortunately it can 
be trivially corrected by defining a rescaled kernel and spectral function,
\[
\overline{K}(\omega,t) = \frac{\omega}{2T} K(\omega,t),
\;\;\;\;\;\;\;\;
\overline{\rho}(\omega) = \frac{2T}{\omega} \rho(\omega),
\]
and performing MEM on
\[
G(t) = \int \frac{d\omega}{2\pi}\, \overline{\rho}(\omega) 
\overline{K}(\omega,t).
\]
We demonstrated that this redefinition removes the unphysical
behaviour of the spectral functions constructed with MEM near $\omega = 
0$, present when the traditional kernel is employed \cite{conductivity}.

In the MEM procedure, we used the default model
\[
\overline{m}(\omega) = m_0 (1 + a\omega)
\]
with $m_0$ determined from a fit of $\int d\omega\, 
\overline{m}(\omega)\overline{K}(\omega)$ to the data.  This definition of 
$\overline{m}(\omega)$ matches the expected perturbative behaviour 
$\rho(\omega) \sim \omega \overline{\rho}(\omega) \sim \omega^2$ as 
$\omega \rightarrow \infty$ and also allows a non-zero intercept in 
$\rho/\omega \sim \overline{\rho}$ (corresponding to a transport peak) as 
$\omega \rightarrow 0$.
For a study of lattice spectral functions at nonzero momentum in the 
infinite temperature limit, see Ref.\ \cite{Aarts:2005hg}.

The lattice action and parameters used are identical to those in Ref.\ 
\cite{conductivity}, i.e.\ a simple Wilson plaquette action with
quenched, staggered fermions.  The lattice parameters used are
displayed in Table~\ref{tb:params}.  In Ref.\ \cite{conductivity} we
determined the electrical conductivity, $\sigma/T = 0.4 \pm 0.1$, a
result which has since been confirmed \cite{karsch}.  We extend the
work in Ref.\ \cite{conductivity} further by using twisted boundary
conditions with the same choices of twists as in Ref.\ \cite{Flynn:2005in}.
This enables us to access a large range of momenta which are listed in
Table~\ref{tab:mom}. Preliminary results can be found in Ref.\ 
\cite{Aarts:2006wt}.


\begin{table}
\begin{center}
\begin{tabular}{lccc}
\hline
&&&\\
                 &                   & {\bf Cold}               & {\bf Hot} \\
&&&\\
\hline
&&&\\
Spatial Volume   & $N_s^3 \times N_t$ & $48^3 \times 24$    & $64^3 \times 24$ \\
Lattice spacings & $a^{-1}$           & $\sim 4$ GeV        & $\sim 10$ GeV \\
$T$              & $1/(aN_t)$         & $T \sim 160 $MeV$ \sim 0.62 T_c$ 
                                                            & $T \sim 420 $MeV$ \sim 1.5 T_c$ \\
Statistics       & $N_{cfg}$          & $\sim 100$          & $\sim 100$ \\
Quark Masses     & $ma$              & $0.01\;\&\;0.05$    & $0.01\;\&\;0.05$ \\
&&&\\
\hline
\end{tabular}
\end{center}
\caption{Lattice parameters used in the simulation. Estimates for the 
lattice spacing and temperature are taken from Ref.\ \cite{Datta:2003ww}.}
\label{tb:params}
\end{table}


Since we use staggered correlators which have contributions from both 
parity partners,
\[
G(t) = \int \frac{d\omega}{2\pi} K(t,\omega)
\left[ \rho(\omega) - (-1)^{t/a} \; \overline{\rho}(\omega) \right],
\]
we are forced to apply MEM to even and odd times separately, and then
reconstruct the physical spectral function from
\[
\rho^{\rm{phys}} = \frac{1}{2}
\left( \rho^{\rm{even}} + \rho^{\rm{odd}} \right).
\]


\begin{table}
\begin{center}
\begin{tabular}{cccc}
\hline
&&&\\
$\vec{p}L$          & $|p|L$                 & Longitudinal & Transverse \\
&&&\\
\hline
&&&\\
$   (0,\;0,\;0) $   & 0                      & - & - \\
$   (2,\;0,\;0) $   & 2                      & $V_1\;$ & $\;V_2\;$ \& $\;V_3$ \\
$ (0,\;\pm\pi,\;0)$ & $\pi$                  & $V_2\;$ & $\;V_1\;$ \& $\;V_3$ \\
$(-2,\;\pm\pi,\;0)$ & $\sqrt{4+\pi^2}\sim3.72$  & -       & $V_3$ \\
$(0,\;\pm2\pi,\;0)$ & $2\pi\sim6.28$            & $V_2\;$ & $\;V_1\;$ \& $\;V_3$ \\
$(2,\;\pm2\pi,\;0)$ & $2\sqrt{1+\pi^2}\sim6.59$ & -       & $V_3$ \\
$(0,\;3\pi,\;0) $   & $3\pi\sim9.42$            & $V_2\;$ & $\;V_1\;$ \& $\;V_3$ \\
$(-2,\;3\pi,\;0)$   & $\sqrt{4+9\pi^2}\sim9.63$ & -       & $V_3$ \\
&&&\\
\hline
\end{tabular}
\end{center}
\caption{A list of the momenta studied showing which components of the
  vector current, $V_i=\overline\psi\gamma_i\psi$, are longitudinal or
  transverse. The momenta components $\vec{p}L$ which are multiples of
  $\pi$ are obtained from the usual Fourier sum, and those $pL$
  components which are integer-valued are obtained from twisted boundary
  conditions.}
\label{tab:mom}
\end{table}




\section{Longitudinal and Transverse Correlators}

\begin{figure}
\begin{center}
\includegraphics[width=0.7\textwidth]{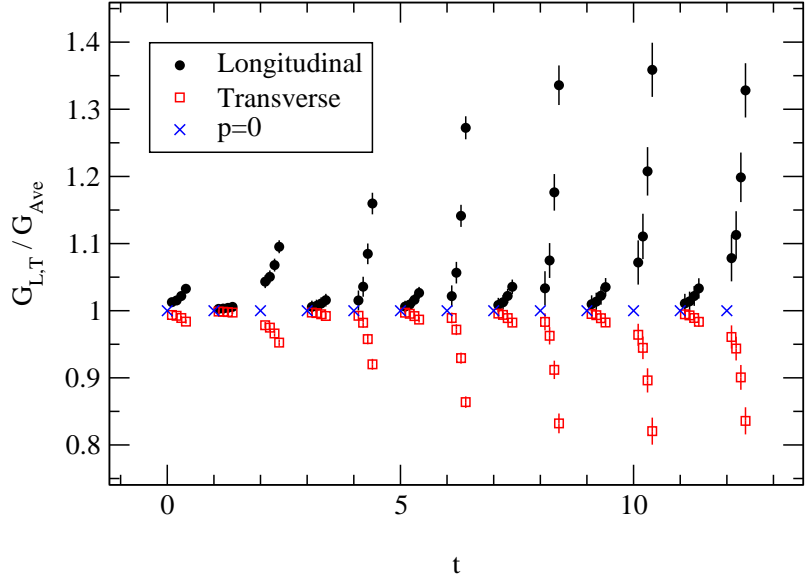}
\caption{ Longitudinal and transverse vector correlation functions
  normalised by the average correlation function, $G_{\rm{Ave}} =
  (G_{\rm{L}} + 2 G_{\rm{T}})/3$, for various momenta, $\vec{p}$,
  as a function of euclidean time.  Data points for each momentum (for a 
  given time) are offset horizontally for clarity; from left to right they
  are $|\vec{p}|L$ = 2, $\pi$, $2\pi$ and $3\pi$.}
\label{fig:longtran}
\end{center}
\end{figure}

In Fig.\ \ref{fig:longtran} we show the vector correlators for both the 
longitudinal and transverse modes, $G_{\rm{L,T}}$, for momenta $pL = 
2,\pi,2\pi,3\pi$ as a function of euclidean time. We display these as a 
ratio with the average correlation, $G_{\rm{Ave}} = (G_{\rm{L}} + 2 
G_{\rm{T}})/3$.
Note that we use local (rather than conserved) currents in our
correlators definitions.
As can be seen, there is a clear difference between these 
modes; the longitudinal correlator is consistently larger than the 
transverse for each time slice and momentum value. 
It would be interesting to understand this feature analytically.
Note that the momentum dependence is much stronger for even
timeslices. This is an artefact of the staggered formulation.

\begin{figure}
  \includegraphics[width=.49\textwidth]{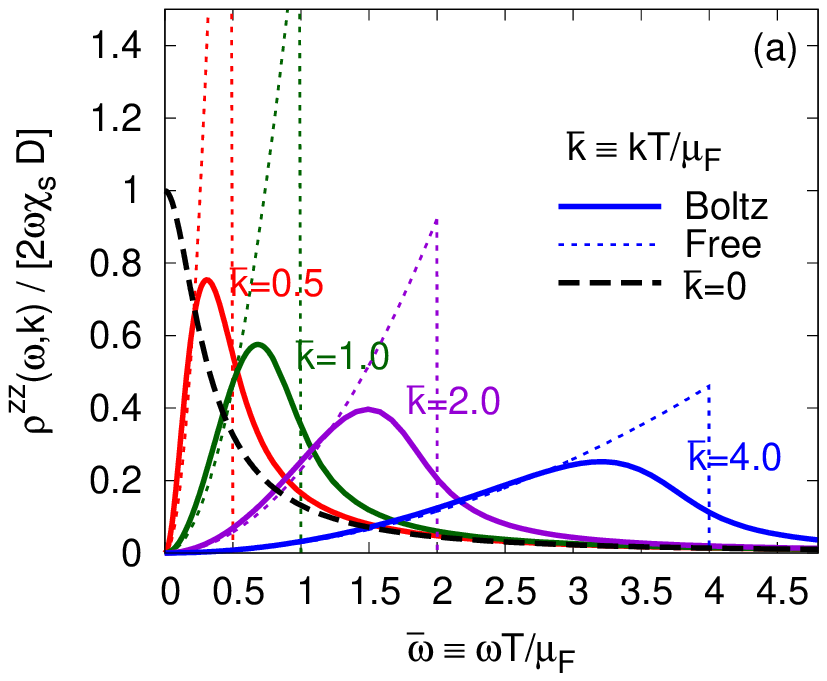}
  \includegraphics[width=.49\textwidth]{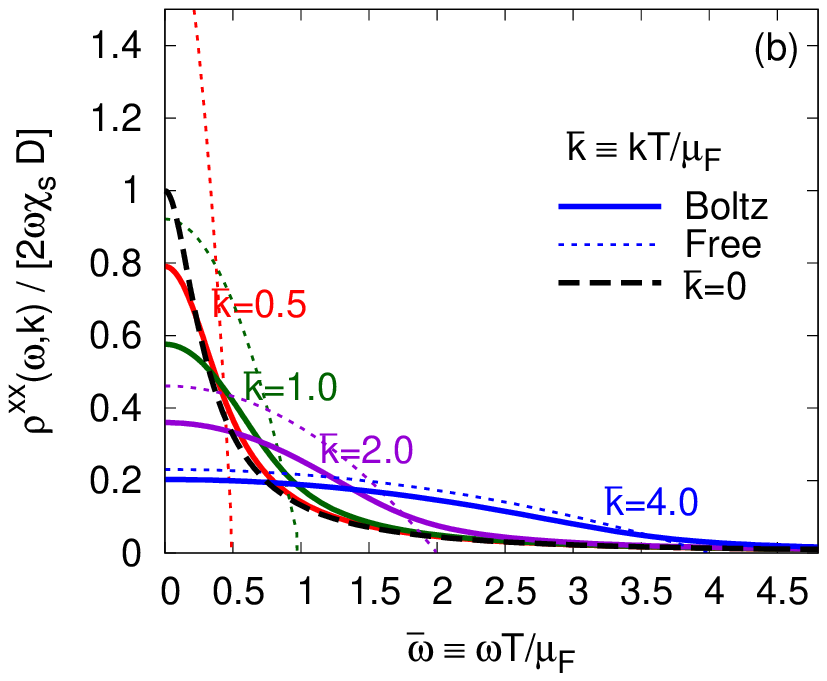}
  \caption{Longitudinal (left) and transverse (right) vector spectral
    function predictions from Ref.\ \cite{Hong:2010at}.}
\label{fig:teaney}
\end{figure}

The diffusivity, $D$, can in principle be obtained from the momentum
dependence of the vector spectral function (at small mass).
In Ref.\ \cite{Hong:2010at} a prediction was made of the energy 
dependence of the
vector spectral function for both the longitudinal and transverse
modes (for the massless case in the plasma phase). This is shown in
Fig.\ \ref{fig:teaney}.  As can be seen, their prediction for the
$\omega \rightarrow 0$ behaviour differs between these two modes:
$\rho_{\rm{L}}(\omega=0, \vec{p}) = 0$ for $\vec{p} \ne 0$ whereas
$\rho_{\rm{T}}(\omega=0, \vec{p}) \ne 0$.

\begin{figure}
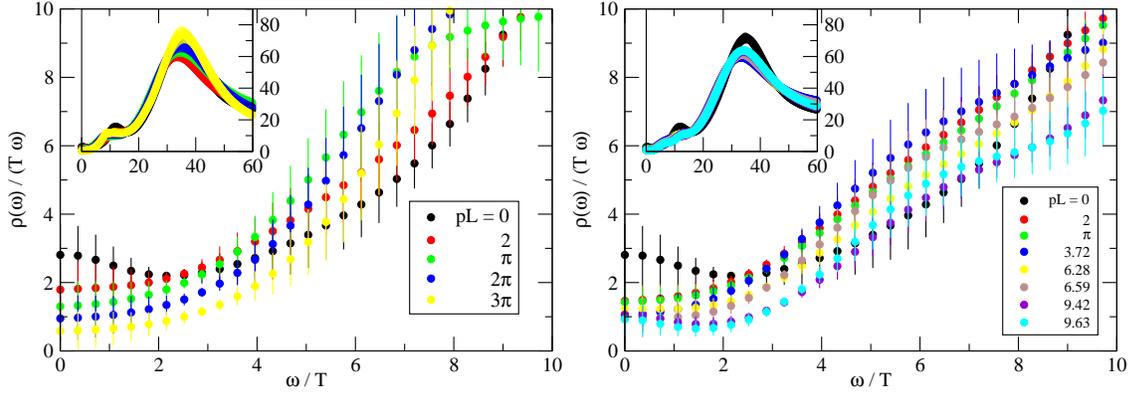

\includegraphics[width=.49\textwidth]{hot_plot_long_m1_both.eps}
\includegraphics[width=.49\textwidth]{hot_plot_tran_m1_both.eps}
\caption{Longitudinal (left) and transverse (right) vector spectral
  function for the light ($ma=0.01$), hot case. The main plot shows
  the $\omega \sim 0$ region, and the insert the full $\omega$ range.}
\label{fig:Vhotlight}
\end{figure}

\begin{figure}
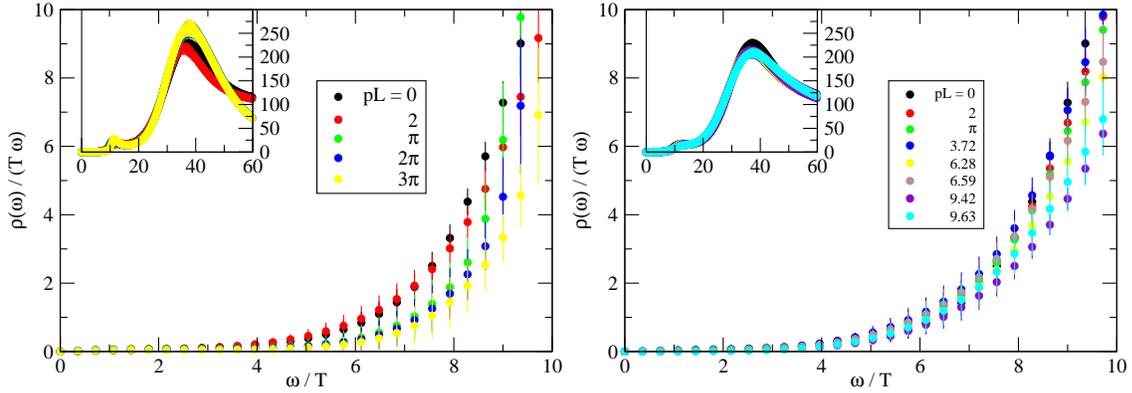

\includegraphics[width=.49\textwidth]{hot_plot_long_m2_both.eps}
\includegraphics[width=.49\textwidth]{hot_plot_tran_m2_both.eps}
\caption{Longitudinal (left) and transverse (right) vector spectral
  function for the heavy, hot case. The main plot shows
  the $\omega \sim 0$ region, and the insert the full $\omega$ range.}
\label{fig:Vhotheavy}.
\end{figure}

In Fig.\ \ref{fig:Vhotlight}, we show the spectral functions obtained
via MEM for the longitudinal and transverse vector case for quark mass
$ma=0.01$. As can be seen, there is a clear non-zero intercept for
both the longitudinal (contradicting \cite{Hong:2010at}) and transverse
modes.  The spectral function for the heavier quark mass, $ma=0.05$,
is shown in Fig.\ \ref{fig:Vhotheavy}.  We note that the intercept is
zero for both the longitudinal and transverse cases for this heavier
quark mass yielding a null result for the heavy-quark diffusion
coefficient \cite{Petreczky:2005nh}.

We checked the spectral function for both the quark masses $ma=0.01$
and 0.05 in the {\em cold} phase, and found a zero intercept for
both longitudinal and transverse modes. This agrees with our
expectation that there are no transport features in this phase.

\begin{figure}
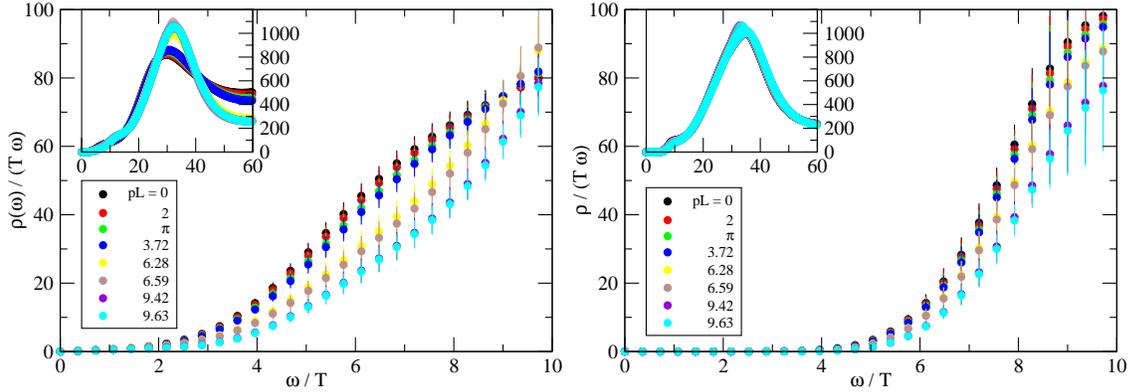

\includegraphics[width=.49\textwidth]{hot_plot_PP_m1_both.eps}
\includegraphics[width=.49\textwidth]{hot_plot_PP_m2_both.eps}
\caption{Pseudoscalar spectral function for the $ma=0.01$ (left) and
  $ma=0.05$ (right) cases. The main plot shows
  the $\omega \sim 0$ region, and the insert the full $\omega$ range.}
\label{fig:PS}
\end{figure}

In Fig.\ \ref{fig:PS}, the pseudoscalar spectral functions are
shown in the hot phase. We observe a zero intercept for this channel
again in agreement with the expectation that there are no transport
features.

\section{Conclusions}

In this work we have extended our previous study \cite{conductivity}
of the spectral functions of vector correlators in the cold and hot
phase of QCD to non-zero momenta using twisted boundary conditions to
allow a finer momentum resolution. We are thus able to calculate and
contrast the longitudinal and transverse modes of the vector two-point
function.  We find that a zero intercept at zero energy for the
spectral function in all cases except for the hot, light quark vector
mesons where we find $\rho(\omega \to 0,\vec{p}) \ne 0$ for both the
longitudinal and transverse modes.  This is despite the fact that the
two modes' two-point correlation functions have a different qualitative
behaviour.  Our future plans are to study the momentum dependence of
these modes with the aim of further studying hydrodynamic structure at 
zero and nonzero momentum \cite{future}.




\end{document}